\documentclass[final,5p,times,twocolumn,numbers,sort]{elsarticle}
\usepackage{amsmath}
\usepackage{amssymb}
\usepackage{lipsum}
\usepackage{nameref}
\usepackage[colorlinks=true,linkcolor=blue,urlcolor=blue,citecolor=blue]{hyperref}

\journal{Physics Letters B}
\begin{document}

\begin{frontmatter}
\title{Noise-induced stabilization of Schwarzschild–AdS black holes under stochastic Ricci flow}
\author[first]{Jihui Wang}
\affiliation[first]{organization={Department of Physics, Fudan University, Shanghai, China},
addressline={2005
Songhu Road}, 
city={Shanghai},
postcode={200438}, 
country={China}}

\author[second]{Matteo Lulli}  
\affiliation[second]{organization={Idea.deploy Labs, 9/F VPOINT, 18 Tang Lung Street},
addressline={Causeway Bay, Hong Kong}, 
city={Hong Kong},
country={China}}

\author[third]{Antonino Marcian\`o}
\affiliation[third]{organization={Department of Physics, Fudan University, Shanghai, China},
addressline={2005
Songhu Road}, 
city={Shanghai},
postcode={200438}, 
state={Shanghai},
country={China}}

\begin{abstract}
We investigate the stochastic Ricci flow of spherically symmetric perturbations of the Schwarzschild--Anti de Sitter black-hole metric. Elaborating on the Ricci-flow analysis of Headrick and Wiseman, we include a negative cosmological constant through a Ricci-target term and study how the flow is correlated with the thermodynamic heat capacity of the black hole. Numerical simulations show that, in the positive heat-capacity regime, perturbations of the angular sector of the metric relax toward the Schwarzschild--Anti de Sitter fixed point, while in the negative heat-capacity regime they grow under the deterministic Ricci flow. We then introduce a multiplicative stochastic noise and find that sufficiently strong stochasticity can suppress the growth of these perturbations, effectively stabilizing configurations that would otherwise be thermodynamically unstable. Finally, we reformulate the dynamics in terms of an entropy variable evolving on a thermodynamic free-energy landscape, and support the metric-flow results through Monte Carlo simulations and the associated Fokker--Planck equation. These results suggest that stochastic fluctuations can modify the relation between geometric stability under Ricci flow and thermodynamic stability in asymptotically Anti de Sitter black-hole spacetimes.
\end{abstract}

\end{frontmatter}

\section{Introduction} \label{introduction}
\noindent 
The geometry gradient (Ricci) flow was originally introduced in \cite{RF5}, being used to describe the flow of $3$-dim Riemannian manifolds. Within this context, a natural approach is to consider the gradient flow of the gravitational action \cite{RF6}. Applying the Ricci flow to $4$-dim Lorentzian manifold, holding more relevance to describe gravity, shows similarity with the stochastic quantization \cite{RF1,SRF2,SQ1}. The analysis developed in \cite{RF1} shows, by assuming the variation of the Einstein-Hilbert action with respect to the metric as the deterministic force term (while a random force comes from the background), that the stochastic version of the Ricci flow is a generalized Langevin equation in the realm of the gravity theory. The random force may arise from quantum effects or thermal fluctuations. Using the ADM formalism, within the Stochastic Ricci Flow (SRF) framework one can recast perturbations to the Schwarzschild metric as a Kardar-Parisi-Zhang (KPZ) equation Ref.~\cite{RF1}. On a different footing, Li and Wang \cite{Li_Wang_1} have proposed that an energy landscape is encoded within stochasticity. They have shown that the Hawking-Page phase-transition happens thanks to fact that enough strength is provided to the Brownian motion, hence triggering the transition from an unstable one to a stable phase. Within this context, the Ricci flow has been studied in order to unveil the thermodynamic properties of the black hole solutions of gravitational theories. Specifically, De Biasio et. al. have studied in Ref~\cite{RF6} the Ricci flow of the Reissner-Nordstr{\"o}m (RN) black hole metric. It turned out that the stability under the geometry gradient flow of the perturbed 2-sphere part within the magnetic RN black hole metric coincides with its heat capacity critical point. The Schwarzschild-anti De Sitter (S-AdS) solution also features a heat capacity critical point. Based on the numerical investigations in \cite{RF2,RF10,RF6}, we have accomplished in this study numerical simulations of the S-AdS metric in presence of stochasticity, also using the thermodynamic energy potential in the stochastic gradient flow, and completed the analysis with numerical simulations.

The addition of a cosmological constant $\Lambda$ to the framework changes the characteristic behavior of the Schwarzschild heat capacity, which is known to be negative. The S-AdS solution exhibits indeed a region of positive heat capacity when $\Lambda$ is sufficiently large in absolute value. In Sec.~\ref{sec:2}, our numerical results show that in the positive heat capacity region --- or equivalently, the positive isothermal compressibility region in the extend phase space --- both quantities change sign at the same point, the angular part of the perturbed S-AdS metric, i.e.  $d\Omega^2=r^2F_3(d\theta^2+\sin^2\theta d\phi^2)$, is stable under the Ricci (target) flow, and finally the perturbation within the function $F_3(\lambda,r)$ dissipate over the flow-time $\lambda$, i.e. $F_3(\lambda,r)\to 1$, when $\lambda\to \infty$. This means the initial perturbed metric evolves toward the S-AdS solution, which is a fixed point of the Ricci target flow equation. Conversely, in the negative heat capacity region --- also corresponding to the negative compressibility region --- the perturbed S-AdS metric evolves divergently. Nonetheless, the stochasticity can still stabilize the perturbation, as shown by the numerical results.  

An important conceptual point of this work is that the thermodynamic description developed in the second part of the paper does not rely on an independent phenomenological construction. Rather, both the Ricci-flow analysis and the entropy-flow description are formulated in the Euclidean sector. The Ricci flow evolves Euclidean Schwarzschild--AdS geometries toward Einstein metrics, while the same Euclidean Einstein solutions dominate the gravitational partition function in the semiclassical approximation. Consequently, the on-shell Euclidean gravitational action naturally provides the thermodynamic free-energy functional governing the effective entropy dynamics. In this sense, the entropy stochastic flow may be regarded as an effective thermodynamic reduction of the underlying stochastic Ricci flow.

We further dig into the correspondence between the Ricci flow and thermodynamic properties of the black hole spacetime. In Sec.~\ref{sec:free energy},  we recast the black hole action as the thermodynamic potential energy in the flow equation, assuming the action $I_G$ to be expressed by $I_G=\beta M-S$. We then evaluate the reference potential energy $I_0$ at the Davies heat capacity divergence point --- i.e., the value of AdS radius $l$ that equates $l_0=\sqrt{3} \, r_H$, $r_H$ denoting the horizon radius --- treating the entropy as the variable under the evolution in the stochastic Ricci flow, $\frac{\delta S}{\delta \lambda}=\frac{\partial (I_{G}-I_0)}{\partial S}+\eta(\lambda)S$. We observe that when the AdS spacetime radius is contained within the horizon, i.e. $l<l_0$, the entropy perturbation $\delta S$ dissipates under the flow evolution, while it blows up when $l>l_0$. This supports the fact that the black hole Einstein-Hilbert action can be effectively related to its thermodynamic free energy.\\

The structure of the paper is summarized as follows. In Sec.~\ref{sec:1}, we review the Ricci target flow and thermodynamic theory of the S-AdS black hole. Sec.~\ref{sec:2} presents the numerical simulation scheme and results. Sec.~\ref{sec:free energy} discusses the stochastic gradient flow of the entropy, based on the thermodynamic free energy. Sec.~\ref{MC} focuses on Monte Carlo and Fokker--Planck analyses. Finally, Sec.~\ref{sec:3} is devoted to provide a summary of our results and to spell out our conclusions. We use the natural units, $G=c=1$.

\section{Schwarzschild--AdS thermodynamics and Ricci-target flow}\label{sec:1}
\noindent 
We may start from the gravitational action \cite{RF1}:
\begin{equation}
    I_{G}=\frac{1}{2\kappa}\int d^4 x \sqrt{-g}R+\int d^4 x \sqrt{-g}\mathcal{L}_{\rm M}\,, \label{eq:action}
\end{equation}
where $\kappa=8\pi$, and $\mathcal{L}_{\rm M}$ denotes the Lagrangian of the matter field. In our theoretical framework, the cosmological constant, which provide the matter source, is negative, and thus the corresponding energy-momentum tensor provides $T_{\mu\nu}=-\frac{\Lambda g_{\mu\nu}}{8\pi}$.

\subsection{Stochastic Ricci flow of S-AdS black holes}
\noindent 
Ricci flow is a geometric flow of the metric. The solution of the Einstein field equation could be viewed as the solution of the Ricci flow, in other words, the fixed point of the flow. The stochastic Ricci target flow can be recast as \cite{SRF2,RF1} :
\begin{equation}
    \frac{\partial g_{\mu\nu}}{\partial \lambda}=\mathcal{G}_{\alpha\beta\mu\nu}\frac{\delta I_{G}}{\delta g_{\alpha\beta}}+\alpha g_{\mu\nu}\eta(\lambda), \label{eq:langevin}
\end{equation}
where $\lambda$ is the (thermal) flow time and $\mathcal{G}_{\alpha\beta\mu\nu}=\frac{2\kappa}{\sqrt{-g}}(g_{\alpha\mu}g_{\beta\nu}+g_{\alpha\nu}g_{\beta\mu}-g_{\alpha\beta}g_{\mu\nu})$ denotes the super-metric. Besides, $\eta(\lambda)$ is a normal distribution followed from the Gaussian distribution,
\begin{equation}
     \langle\eta(\lambda)\eta(\lambda')\rangle=\delta(\lambda-\lambda')\\ \label{eq:noise_pro1}
\end{equation}
\begin{equation}
        dW\thicksim \mathcal{N}(0,d\lambda) \qquad {\rm and} \qquad \eta(\lambda)=\frac{dW}{d\lambda}\,, \label{eq:noise_pro2}
   \end{equation}
where $dW$ is the Wiener process and $\mathcal{N}(0,d\lambda)$ represents the Gaussian distribution. Notice that we assume the noise term is the multiplicative noise which is the noise multiply by the stochastic quantity. \\

It is also important to specify that throughout this work we interpret the stochastic differential equations in the It\^o sense. The multiplicative noise term appearing in Eq.~\eqref{eq:langevin} is therefore treated according to  It\^o calculus, and all numerical simulations are implemented using the corresponding  It\^o discretization. This prescription is also adopted in the derivation of the associated Fokker--Planck equation discussed in Sec.~\ref{Sec:FP}. The use of the It\^o interpretation is natural in the present framework, since the stochastic force is modeled as an external white-noise source acting independently at each flow-time step.\\

Substituting the action Eq.~\eqref{eq:action} into Eq.~\eqref{eq:langevin}, we obtain the flow equation in a more familiar form, namely 
\begin{equation}
\begin{aligned}
    \frac{\partial g_{\mu\nu}}{\partial \lambda}=&-2(R_{\mu\nu}-R^{T}_{\mu\nu})+\alpha g_{\mu\nu}\eta(\lambda)\\
    =&-2(R_{\mu\nu}-\kappa(T_{\mu\nu}-\frac{1}{2}g_{\mu\nu}T))+\alpha g_{\mu\nu}\eta(\lambda)\,,
    \label{eq:Ricci_target_flow}
    \end{aligned}
\end{equation}
where $R^{T}_{\mu\nu}=\kappa \left(T_{\mu\nu}-\frac{1}{2}g_{\mu\nu}T\right)$ denotes the target Ricci tensor induced by the matter field --- accordingly, we can introduce the Ricci target scalar $R^{T}=g^{\alpha\beta}R^{T}_{\alpha\beta}$. 

\subsection{Thermodynamics of S-AdS black holes}\label{Sec:Thermo}
\noindent
Since the Ricci flow implements the gradient flow of the Einstein equations in the vacuum, there have been several attempts in the literature to relate it to black hole thermodynamics \cite{RF4,RF6,RF2}. The thermodynamic properties of the S-AdS black hole solution have continued to attract the researchers' interest since the pioneering work of Hawking and Page appeared in 1980s \cite{thermo6}. Subsequent studies have included detailed analyses concerning thermodynamic potentials \cite{thermo11}, the Ruppeiner thermodynamic geometry method \cite{thermo8}, and  topological classifications \cite{thermo7,thermo11}.\\
The first law of Schwarzschild-AdS thermodynamics reads
\begin{equation}
    dM=TdS\,,
\end{equation}
where $M$ is the black hole mass and $S$ denotes the entropy. Another expression for the first law of black hole thermodynamics in the so-called extended phase space \cite{thermo5} is provided by the relation
\begin{equation}
    dM=TdS-PdV\,,
\end{equation}
where $P$ is related to the cosmological constant $\Lambda$ ---  we can write the cosmological constant as $\Lambda=-3/l^2$, with $l$ radius of the AdS space --- as  
\begin{equation}
    P=3/(8 \pi l^2)\,,
\end{equation}
$V$ denoting the conjugate quantity that encodes the volume within the horizon $V=\frac{4}{3}\pi r^3$.
The general form of the S-AdS metric can be written as 
\begin{equation}
\begin{aligned}
    ds^2
    =&-f(r)dt^2+f(r)^{-1}dr^2+r^2d\Omega^2\\
    =&-(1-\frac{2M}{r}-\frac{\Lambda r^2}{3})dt^2+(1-\frac{2M}{r}-\frac{\Lambda r^2}{3})^{-1}dr^2\\
    &+r^2(d\theta^2+sin^2\theta d\phi^2)\,.
    \end{aligned} \label{eq:metricg}
\end{equation}
where the $d\Omega^2=r^2(d\theta^2+sin^2\theta d\phi^2)$ is the angular part of the metric, $M$ and $\Lambda$ are the mass of the black hole and cosmological constant. The S-AdS solution is equipped with one event horizon, the horizon radius $r_{H}$ fulfilling $f(r_{H})=0$.
With the metric form in Eq.~\eqref{eq:metricg}, we can proceed to compute related thermodynamic quantity, including the entropy $S$, the temperature $T$ and the specific heat capacity $C_{P}$. Respectively, the entropy can be expressed \cite{Bekenstein,SAdS_ref2} as  
\begin{equation}
    S=\frac{A}{4}=\frac{1}{4}\int \!\!\sqrt{g_{\theta\theta}g_{\phi\phi}}\,d\theta \, d\phi=\pi \,{r_{H}}^{2}\,, \label{eq:area}
\end{equation}
where $A$ represents the area of the horizon, while the Hawking temperature, proportional to the surface gravity \cite{Hawking_AdS}, can be written as \cite{SAdS_ref2} 
\begin{equation}
    T={{\frac{f^{'}(r)}{4\pi}} \Big|}_{r=r_{H}}=\frac{3 r_{H}^2+l^2}{4\pi r_{H}l^2}\,.
\end{equation}
One can then find that
\begin{equation}
\begin{aligned}
    \frac{\partial T(r_{H},l)}{\partial S(r_{H})}&=\frac{1}{2\pi r_{H}}\frac{\partial T}{\partial r_{H}}=\frac{1}{8\pi^2 r_{H}}\left(\frac{3}{l^2}-\frac{1}{{r_{H}}^2}\right)\,,
    \end{aligned}
\end{equation}
from which it follows that
\begin{equation}
    C_{P}=T{\left(\frac{\partial S}{\partial T}\right)}_{P}=2 \pi {r_{H}}^2\frac{3{r_{H}}^2+l^2}{3r_{H}^2-l^2} \,.
\label{eq:heatc}
\end{equation}
\begin{figure}
	\centering 
	\includegraphics[width=0.45\textwidth, angle=0]{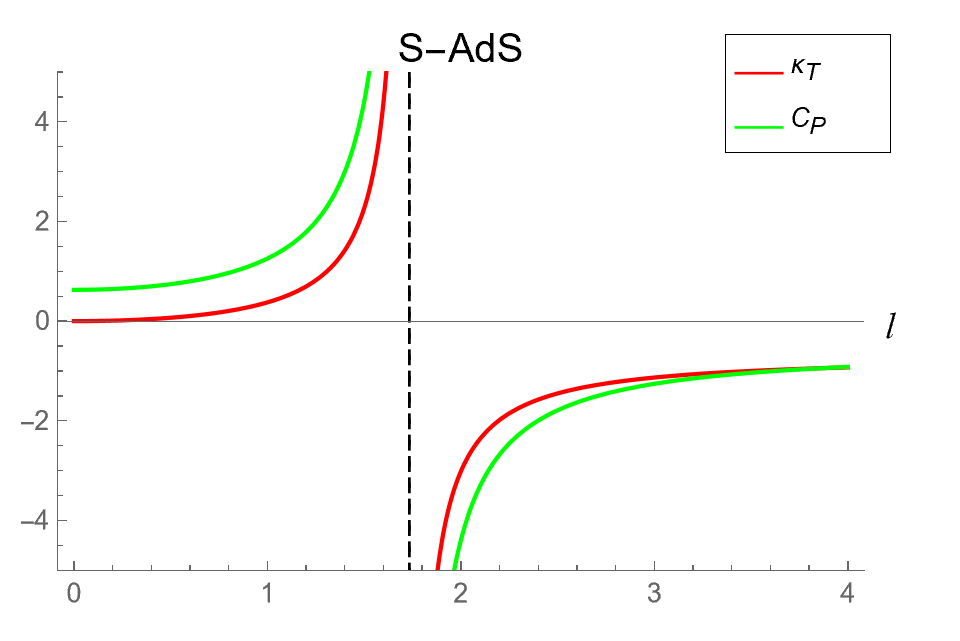}
	\caption{We dispaly the heat capacity and the compressibility of the $S-AdS$ metric, having set $r_{H}=1$. The horizontal axis represents the AdS radius $l$. We display the result multiplying $\kappa_T$ by a coefficient $0.01$, and $C_{P}$ by $0.1$, to achieve clearer visualization.}
	\label{fig_1}%
\end{figure}
Due to the term $PdV$ in the first law of thermodynamics, one can consider the isothermal compressibility $\kappa_{T}$ as a thermal response function~\cite{thermo2}, hence finding   
\begin{equation}
\begin{aligned}
    \kappa_{T}=&-\frac{1}{V}{\left(\frac{\partial V}{\partial P}\right)}_{T}=-\frac{3}{r_{H}}\frac{1}{{\left(\partial P/\partial{r_{H}}\right)}_{T}}=-\frac{3}{r_{H}}\frac{1}{\frac{\partial P}{\partial{T}}\frac{\partial T}{\partial{r_{H}}}}\\
    =&\frac{24\pi l^2 {r_{H}}^2}{3{r_{H}}^2-l^2}\,, \label{eq:compress}
\end{aligned}
\end{equation}
consistent with the expression of the heat capacity in Eq.~\eqref{eq:heatc}.\\ 

We observe from Fig.~\ref{fig_1} that the heat capacity changes sign at $l_0=\sqrt{3}\, r_{H}$. This point coincides with the divergence of the heat capacity and the minimum of the Hawking temperature. Following the literature --- see \cite{thermo11} and Sec.~6.5 in Ref.~\cite{thermo3} --- we refer to it as the Davies point. Although it separates branches with positive and negative heat capacity, its interpretation as a genuine thermodynamic phase transition remains more subtle than the Hawking--Page transition, which corresponds to a change of the globally preferred thermodynamic state. In the present work we therefore regard the Davies point primarily as a stability threshold separating thermodynamically stable and unstable branches. The isothermal compressibility $\kappa_{T}$, as defined in Eq.~\eqref{eq:compress}, then changes sign at the same point where the heat capacity $C_{P}$ changes sign. Therefore, in the case of S-AdS black hole, the inclusion of $\kappa_{T}$ does not impose additional constraints about the thermal stability. \\

Two characteristic thermodynamic phenomena occur in Schwartzschild-AdS black holes. The first is the Hawking--Page transition, which corresponds to a first-order transition between thermal AdS and a large black-hole phase and is related to the minimum of the free energy~\cite{Hawking_AdS}. The second is the Davies point, at which the heat capacity diverges and changes sign. The latter separates a branch with positive heat capacity from a branch with negative heat capacity and therefore marks a change in local thermodynamic stability. Throughout this paper we focus on this stability threshold and its relation to the Ricci-flow dynamics.

\subsection{The metric formalism}
\noindent 
In order to proceed to the numerical simulation, we need a more suitable metric formalism. Thus we recast the metric function $f(r)$ in Eq.~\eqref{eq:metricg} according to
\begin{equation}
\begin{aligned}
    f(r)=&\frac{r^2}{3a_1 a_2}\left(1-\frac{a_1-a_2}{r}\right)\times\\
    &\left(1+\frac{a_1-a_2-i\sqrt{3}(a_1+a_2)}{2r}\right)\left(1+\frac{a_1-a_2+i\sqrt{3}(a_1+a_2)}{2r}\right)\,,
\end{aligned}
\end{equation}
where $a_1$ and $a_2$ are parameters fulfilling 
\begin{equation}
    \begin{aligned}
        \Lambda=-\frac{1}{a_1 a_2}\,, \;
         l^2=3a_1 a_2\,,
       \;  2M=\frac{(a_1-a_2)(a_1^2+a_1a_2+a_2^2)}{3a_1 a_2}\,.
    \end{aligned} \label{eq:relation}
\end{equation}
As there exist $1$ real positive root and $2$ complex roots of $f(r)=0$, which are exactly 
\begin{equation}
        r_{1}=a_1-a_2\,,
\end{equation}
\begin{equation}
        r_{2}=-\frac{a_1-a_2}{2}+i\frac{\sqrt{3}(a_1+a_2)}{2}\,,
\end{equation}
\begin{equation}
        r_{3}=-\frac{a_1-a_2}{2}-i\frac{\sqrt{3}(a_1+a_2)}{2}\,,
\end{equation}
the two parameters should satisfy the condition $a_1>a_2$ in order to preserve the positivity of $M$, the mass of the black hole. The real root $r_1$ individuates the radius of the event horizon, while the conjugate complex roots $r_{2,3}$ can be thought as unphysical apparent horizons~\cite{SAdS_ref1}. As $\Lambda\to0$, the real part of $r_{2,3}$ approaches $-\frac{r_1}{2}$, which is negative, while its imaginary part approaches infinity. Nonetheless, these complex roots may play a role in the Lorentzian version of the stochastic Ricci flow.

We also perform the radial coordinate transformation $r=\frac{1}{1-\rho^2}$, from which, by setting $a_1-a_2=1$, we derive
\begin{equation}
\begin{aligned}
        f(r)=&\frac{1}{3a_1 (a_1-1)}\frac{\rho^2}{(1-\rho^2)^2}\times(1+\frac{1-i\sqrt{3}(2a_1-1)}{2}(1-\rho^2))\\&\times(1+\frac{1+i\sqrt{3}(2a_1-1)}{2}(1-\rho^2))
\end{aligned}\,. \label{eq:metricn}
\end{equation}

\section{Numerical evolution of metric perturbations}\label{sec:2}
\subsection{Settings}
\noindent
Starting from Eq.~\eqref{eq:metricn} and Eq.~\eqref{eq:metricg}, we assume that the metric keeps the form
\begin{equation}
\begin{aligned}
    ds^2&=-\frac{1}{3a_1 a_2}\frac{\rho^2}{(1-\rho^2)^2}(1+\frac{1+i \sqrt{3}(a_1+a_2)}{2}(1-\rho^2))\times\\
    &(1+\frac{1-i \sqrt{3}(a_1+a_2)}{2}(1-\rho^2))F_1(\tau,\rho)dt^2\\
    &+\frac{12a_1 a_2}{(1-\rho^2)^2}\frac{1}{(1+\frac{1+i \sqrt{3}(a_1+a_2)}{2}(1-\rho^2))}\\\times
    &\frac{1}{(1+\frac{1-i \sqrt{3}(a_1+a_2)}{2}(1-\rho^2))}F_2(\tau,\rho)d\rho^2\\
    &+\frac{1}{(1-\rho^2)^2}F_3(\tau,\rho)(d\theta^2+sin^2\theta d\phi^2)
\end{aligned} \,. \label{eq:metricansatz}
\end{equation}
We use this form of metric within the Ricci target flow equation, Eq.~\eqref{eq:Ricci_target_flow}. Besides that, we also include DeTurck term \cite{RF3,RF6} $Lie_{\xi}g_{\mu\nu}=\nabla_{\mu}\xi_{\nu}+\nabla_{\nu}\xi_{\mu}$ in the RHS of the equation --- in order to enhance the numerical stability --- with $\xi^{\lambda}=g^{\alpha\beta}\left(\Gamma^{\lambda}_{\;\alpha\beta}\left(g\right)-\bar{\Gamma}^{\lambda}_{\;\alpha\beta}\left(\bar{g}\right)\right)$.\\

For the choice of the reference metric $\bar{g}_{\mu\nu}$ in the DeTurck term, we adopt the S-AdS metric solution. This implies that the reference metric functions $\bar{F}_i(\tau, \rho)$ take constant values, namely $\bar{F}_1 = \bar{F}_2 = \bar{F}_3 = 1$. The Ricci tensor $R_{\mu\nu}$ can then be calculated through the xAct package for the Mathematica software \cite{xact}. Furthermore, we use $2nd$ order central finite difference method. The range of the radial coordinate is $\rho \in \left(0,1\right)$, with $N=25$ discretization points $\rho_{i}, i=1,\dots,N$. At the boundaries, we impose Neumann boundary conditions for the metric functions, i.e. $\partial F_{i}(\tau,\rho)/\partial \rho=0$ both at the horizon $\rho_0=0$ and at the AdS boundary $\rho_{N+1}=1$. 
Using $2nd$ order forward and backward finite difference formulas to explicitly impose the Neumann boundary conditions, we can write
\begin{equation}
    \begin{aligned}
        F_1(\lambda,\rho_0)&=\frac{1}{4}(3F_1(\lambda,\rho_1)-F_1(\lambda,\rho_2)) \,,\\
        F_1(\lambda,\rho_{N+1})&=\frac{1}{4}(3F_1(\lambda,\rho_{N})-F_1(\lambda,\rho_{N-1}))\,,
    \end{aligned}
\end{equation}
the same formulas are also applied to the other two metric components functions $F_2$ and $F_3$ at the boundaries. The choice of time step length is according to the Courant-Friedrichs-Lewy (CFL) numerical stability condition~\cite{CFL} of finite difference method $\delta \lambda \leq 0.5\times(\delta \rho)^2$, with the equal sign.\\

The initial metric value, i.e., the metric configuration at the flow time $\lambda=0$, are set to be 
\begin{equation}
\begin{aligned}
    &F_1(\lambda=0,\rho)=1\,,\\
    &F_2(\lambda=0,\rho)=1\,,\\
    &F_3(\lambda=0,\rho)=1+0.2\times(1-\rho^2)^2\,,
\end{aligned}
\end{equation}
where the function $F_3(\lambda=0,\rho)$ includes increased perturbations. According to Eq.~\eqref{eq:area}, we can recover the black hole entropy $S$ as a function of the flow time $\lambda$, i.e.
\begin{equation}
    S=\pi F_3\left(\lambda,\rho=0\right){r_{H}}^2\,. \label{eq:F3}
\end{equation}
Based on Eq.~\eqref{eq:F3}, we can use $\bar{S}\equiv\frac{S}{\pi}=F_3(\lambda,\rho=0)\,{r_{H}}^2$ to effectively represent the change of the entropy $S$ as a function of the flow time $\lambda$. This is shown in Sec.~\ref{subsec:2}, where we display the metric function evolution in order to inquire whether this can reach equilibrium or its perturbations grow as $\lambda\to\infty$.

\subsection{The numerical results}\label{subsec:2}
\noindent 
We investigate the evolution, under the stochastic Ricci flow and for different values of $\Lambda$, of the angular components of the perturbed S-AdS metric. As in Eq.~\eqref{eq:F3}, the spherical function $F_{3}$ is related to the horizon radius and to the entropy in the perturbed S-AdS metric. In order to investigate the Davies point, we start focusing on those cases that correspond to the values of the cosmological constant $\Lambda=-1.33$ and $\Lambda=-0.5$, namely the two phases of the heat capacity.\\

Specifically, for the functions that appear in the metric components --- for instance, for $F_3(\lambda)$ --- the corresponding stochastic differential equations can be obtained by applying It\^o's lemma
\begin{equation}
\begin{aligned}
        \frac{\partial F_3(\lambda)}{\partial \lambda}=&\frac{\partial F_3}{\partial g_{22}}\frac{\partial g_{22}}{\partial \lambda}+\alpha^2 g^2_{22}\frac{\partial ^2 F_3}{\partial g_{22}^2}\\
    =&\frac{\partial F_3}{\partial g_{22}}(-2R_{22}+\alpha g_{22} \eta(\lambda))
\\=&-\frac{2}{r^2}R_{22}+\alpha F_3(\lambda) \eta(\lambda)\,,
\end{aligned}
\end{equation}
and similarly for other $F_i$, i.e. 
\begin{equation}
\begin{aligned}
        \frac{\partial F_1(\lambda)}{\partial \lambda}=&\frac{\partial F_1}{\partial g_{tt}}\frac{\partial g_{22tt}}{\partial \lambda}+\alpha^2 g^2_{tt}\frac{\partial ^2 F_3}{\partial g_{tt}^2}\\
    =&\frac{\partial F_1}{\partial g_{tt}}\left(-2R_{tt}+\alpha g_{tt} \eta(\lambda) \right)\,.
\end{aligned}
\end{equation}

\vspace{0.2 cm}

\begin{figure}
	\centering 	\includegraphics[width=0.45\textwidth, angle=0]{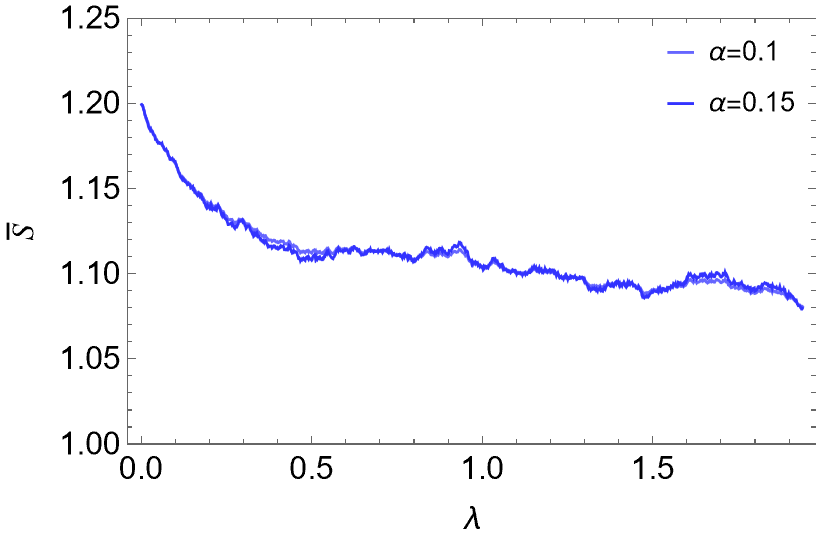}
	\caption{For out-of-equilibrium perturbations of the $S-AdS$ metric, for which $\Lambda=-1.33$, we display the evolutions of $\bar{S}$ with the thermal time $\lambda$. Two curves are shown, representing the evolution under two different values of the stochasticity strength, $\alpha=0.1,0.15$. } 
	\label{fig_M1}
\end{figure}

In Fig.~\ref{fig_M1}, we first provide the results  setting the parameters $a_{1}=3/2$, $a_{2}=1/2$, the cosmological constant to take the value $\Lambda=-1.33$, and the AdS radius square to be $l^2=2.25$, according to Eq.~\eqref{eq:relation}. We add the stochasticity in the strengths $\alpha=0.1$ and $0.15$. We run 100 stochastic trajectories and only show the mean trajectory, which is averaged over these different realizations. 
\begin{figure}[htbp]
\centering
\includegraphics[width=.45\textwidth]{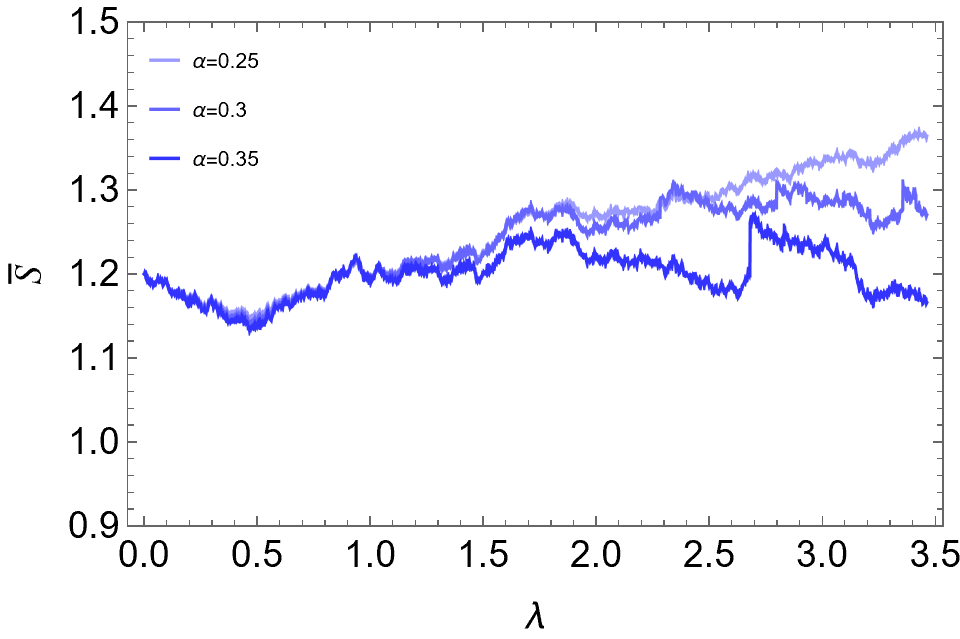}
\caption{For out-of-equilibrium perturbations of the $S-AdS$ metric, for which $\Lambda=-0.5$, $l^2=6$. We display the evolution of $\bar{S}$ with the thermal time $\lambda$.  From the top to bottom corresponds to evolutions with different stochasticity strengths $\alpha=0.25,0.3,0.35$.}\label{fig_M2}
\end{figure}
We further observe from Fig.~\ref{fig_M1} that initial perturbations of $F_3(\tau=0,\rho)$, corresponding to perturbations of the event horizon ${r_{H}}^2$, actually tend to converge with the flow time $\lambda\to \infty$. This indicates that the perturbed metric converges under the flow toward the S-AdS solution when the AdS radius $l<l_0$, this with a large cosmological constant, the black hole horizon is stable against the perturbation. \\

In Fig.~\ref{fig_M2}, we then present the results obtained for a choice of the parameters corresponding to $a_1=2$ and $a_2=1$, which is consistent with $l^2=6$ and hence $\Lambda=-0.5$. Fig.~\ref{fig_M2} illustrates the evolution under the flow of the perturbed metric function $\bar{S}=F_{3}(\rho=0,\lambda)$. As discussed in Sect.~\ref{sec:2}, $l=\sqrt{6}$ is larger than the critical value $l=\sqrt{3}$. When the AdS radius exceeds the critical value set by the horizon scale, the heat capacity becomes negative, individuating a thermodynamical instability. We observe that the increasingly perturbed metric function $\bar{S}$ expands when the amplitude of the stochastic noise is small, whereas for larger stochasticity strength, i.e. $\alpha> 0.3$, it becomes stagnant and decreases slightly with the flow of the thermal time. Roughly, from the results of the numerical simulation  we can estimate the critical value of the stochasticity strength to be $\alpha_0=0.3$. Beyond this threshold, stability can be attained.

Summarizing, we have learned that when the AdS radius $l$ acquires values below or above the critical point, the thermal time dynamics dictated by the Ricci target flow (of the perturbed components of the metric function $F_{3}(\lambda,\rho=0)$) entails very different behaviors. The metric configurations exhibiting negative heat capacity or negative compressibility do not converge toward the S-AdS fixed point, but instead evolve toward divergent configurations. However, under strong stochasticity, the S-AdS solution becomes thermodynamic stable even with a negative heat capacity. 

\begin{figure}[htbp]
\centering
\includegraphics[width=.45\textwidth]{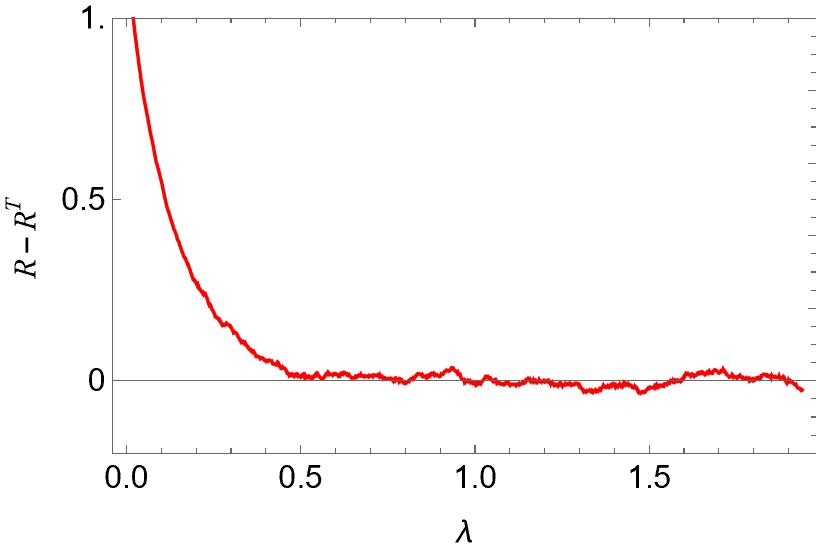}
\caption{For out-of-equilibrium perturbations of the $S-AdS$ metric, for which $\Lambda=-1.33$, $l^2=2.25$, $\alpha=0.1$, we display the evolution in the thermal time $\lambda$ of $R-R^{T}$.}\label{fig:R1}
\end{figure}

\begin{figure}[htbp]
\centering
\includegraphics[width=.45\textwidth]{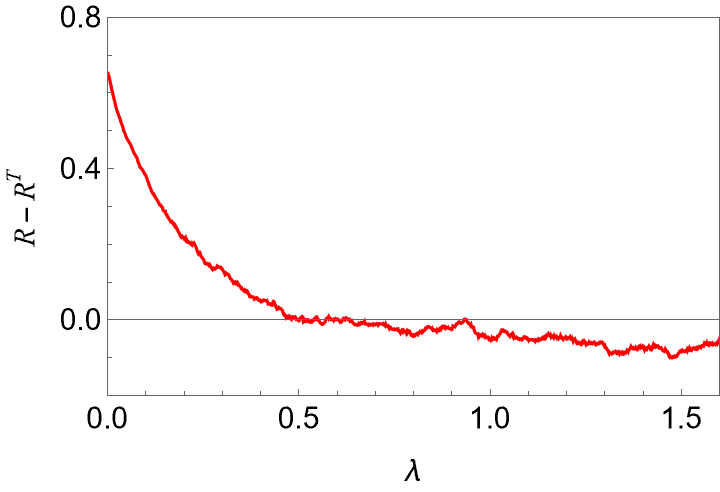}
\caption{For out-of-equilibrium perturbations of the $S-AdS$ metric, for which $\Lambda=-0.5$, $l^2=6$, $\alpha=0.3$, we display the evolution in the thermal time $\lambda$ of $R-R^{T}$.}\label{fig:R2}
\end{figure}

\vspace{0.2 cm}

We have further inspected the evolution of the effective Ricci scalar $R-R^{T}$ in Fig.~\ref{fig:R1} and in Fig.~\ref{fig:R2}. The fact that the effective Ricci scalar decreases with the flow/thermal time $\lambda$ in both cases, indicates that it approaches the solution of the Einstein field equation. In the case $l^2=6$, the $R-R^T$ does not stabilize around zero, but instead continues decreasing toward negative values. This indicates the perturbations of a spherically-symmetric metric may deviate from the solutions of the Einstein equations, and thus the entropy will not converge to the equilibrium value. 

\section{Entropy-flow model and thermodynamic potential}\label{sec:free energy}
\noindent 
In the following we adopt the Euclidean semiclassical formulation of black-hole thermodynamics developed by Gibbons and Hawking \cite{thermo9} 
and later exploited by Witten \cite{Witten1998} in the context of Schwarzschild--AdS black holes. Within this framework, the gravitational path integral is dominated by the classical Euclidean saddle, and the on-shell Euclidean gravitational action coincides with the thermodynamic action,
\begin{equation} \label{RoBER}
I_G=I_E=\beta F\,,
\end{equation}
where $F$ denotes the Gibbs free energy of the black-hole configuration. Consequently, once evaluated on the Schwarzschild--AdS solution, the gravitational action may be regarded as an effective thermodynamic potential governing the entropy dynamics. The stochastic gradient flow introduced below should therefore be understood as an effective reduced description of the full stochastic Ricci-flow dynamics. 
%
%

This treatment is more akin to the free energy landscape and is related to the generalized free energy method \cite{Li_Wang_1,Wei_1}. Within this framework, we then aim to show how the Langevin-type equation that is based on the thermodynamic free energy can be exploited to study the problem of the divergence of the black hole thermodynamic heat capacity.

\subsection{Derivation of the free energy from the action}
\noindent
In Ref.~\cite{Li_Wang_1}, the authors systematically use thermodynamic functions related to the action and employ the stochastic Langevin equation of the black hole thermodynamic free energy in order to investigate the Hawking--Page phase transition, in which the Gibbs free energy appears as the potential and the horizon radius $r_h$ as the order parameter, constructing a Langevin equation for the parameter $r_h$. 
This method is intimately related to the stochastic Ricci flow, as both are based on the Einstein-Hilbert action. In this paper, we focus on the heat capacity divergence point and entropy perturbation stability under the stochastic gradient flow in Eq.~\eqref{eq:langevin}, and propose a simplified stochastic gradient flow ---
notice that according to Hawking and Page \cite{thermo6}, $I_{G}$ can be expressed as function of $S$ --- for the entropy $S$,
\begin{equation}
    \frac{\partial S(\lambda)}{\partial \lambda}=\frac{\delta I_{G}}{\delta S}+S(\lambda)\, \eta(\lambda)\label{eq:free energy Ricci flow}
\end{equation}
where $S=A_{\rm area}/4=\pi r_{H}^2$ is the entropy of the Kerr-Newman black hole and the time parameter is denoted with $\lambda$. The Gaussian white noise $\eta(\lambda)$ obeys Eq.~\eqref{eq:noise_pro1} and Eq.~\eqref{eq:noise_pro2}. This equation aims at qualitatively representing the same result of the stochastic Ricci flow of the metric discussed within Sec.~\ref{sec:2}.
In the following, we will briefly justify this equation and write it explicitly for the Schwarzschild-AdS metric. 

The identification of the on-shell Euclidean gravitational action with the thermodynamic action follows from the Euclidean formulation of black-hole thermodynamics developed by Gibbons and Hawking \cite{thermo9} and from its application to Schwarzschild--AdS black holes by Witten \cite{Witten1998}. In particular, for spherically symmetric spacetimes Eq.~\eqref{RoBER} can be derived directly within the horizon thermodynamics approach developed by Padmanabhan \cite{Paddy_2}. 
%
%
%
%
%
%
Thus, once evaluated on the Euclidean Schwarzschild--AdS saddle, the gravitational action is exactly the thermodynamic action $\beta F$, allowing the entropy to be treated as an effective collective variable evolving on the corresponding free-energy landscape. 

In the following we do not identify the local Einstein--Hilbert functional in Eq.~\eqref{eq:action} directly with a function of the entropy. Rather, we introduce an effective thermodynamic action, denoted by $I_{\rm th}(S,l)$, obtained from the Euclidean gravitational thermodynamics of the Schwarzschild--AdS black hole. In the semiclassical Euclidean approach, the on-shell gravitational action determines the canonical partition function and satisfies $I_{th}=\beta F$, where $F=U-TS$ is the thermodynamic free energy, $U=M$ denoting the internal energy, and $M$ mass of the black hole. Therefore, once the AdS radius $l$ is fixed, $I_{th}$ may be regarded as an effective potential for the entropy variable $S$. This construction is not meant to replace the metric Ricci-flow dynamics, but to provide a reduced thermodynamic description that captures the stability properties observed in the full metric-flow simulations.

\begin{figure}
	\centering 
\includegraphics[width=0.45\textwidth, angle=0]{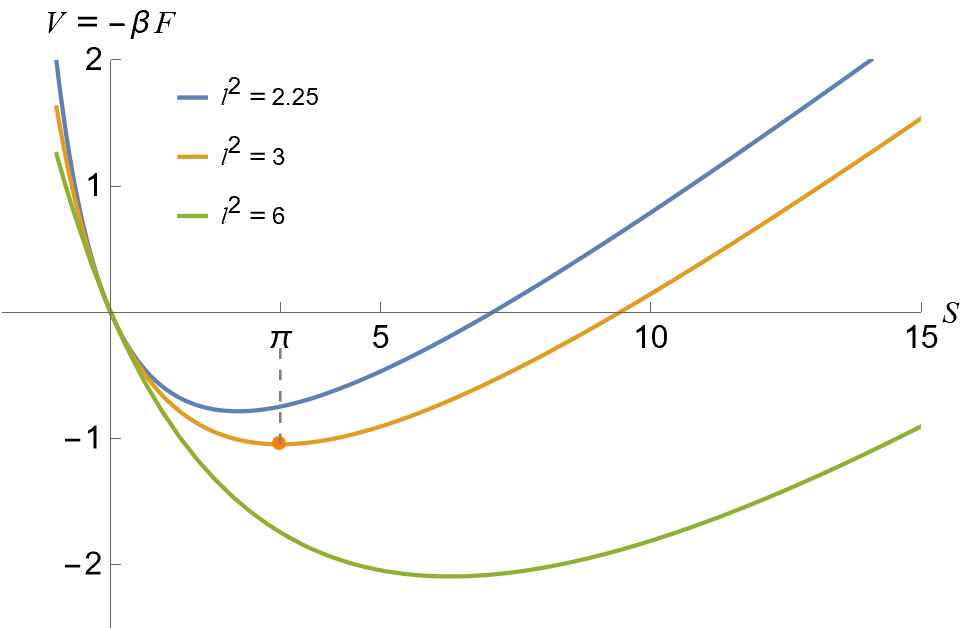}
	\caption{We display the potential entering the stochastic gradient flow as a function of the entropy $S$, the square of the AdS radius taking the value $l^2=2.25$, $l^2=3$ and $l^2=6$.}
\label{fig_V1}
\end{figure}

Relying on these observation, we calculate the related quantities for the S-AdS spacetime, including the mass $M$ and Hawking temperature $T$, expressed as the function of the black hole entropy $S$, namely 
\begin{equation}
    M=\frac{1}{2}\sqrt{\frac{S}{\pi}}\left(1+\frac{S}{\pi l^2}\right),
\end{equation}
\begin{equation}
    T=\frac{\partial M}{\partial S}=\frac{1}{4\sqrt{\pi S}}\left(1+\frac{3S}{\pi l^2}\right),
\end{equation}
and 
\begin{equation}
    F=T S\cdot\frac{\pi l^2 -S}{\pi l^2 +3 S}. \label{eq:Free energy}
\end{equation}
We identify the $-\beta F$ as the potential $V$ of the thermodynamic system, namely 
\begin{equation}
    V(S,l)=-I_{th}(S,l)=-\beta M+S=-S\frac{\pi l^2 -S}{\pi l^2 +3 S}\label{eq:thermodynamic potential}\,.
\end{equation}
Once the value of $l$ is fixed, $V(S,l)$ only depends on $S$. In the following, we then plot the thermodynamic potential $V$ as a function of $S$, according to Eq.~\eqref{eq:Free energy}. In Fig.~\ref{fig_V1} we plot $V$ as a function $S$ for  $l^2=2.25$, $l^2=3$ and $l^2=6$.
We remind that the S-AdS black hole spacetime has an event horizon located at $r=r_H$, and that the cosmological constant $\Lambda$ is related to the AdS radius $l$ by $\Lambda=-3/l^2$.  From Fig.~\ref{fig_V1} we observe that when the AdS radius takes the critical value $l_0=\sqrt{3} r_H$, at which the divergence of the heat capacity occurs, the potential $V(S)$ exhibits a local minimum for which the entropy $S$ equates the black hole entropy $S_{BH}=\pi r_H^2=\pi$. This indicates that, when $l=l_0$, the equilibrium point of the thermodynamic potential $V(S)$ is located at the event horizon entropy $S=S_{BH}$.\\

Motivated by this exact thermodynamic interpretation of the Euclidean gravitational action, we consider the effective stochastic gradient flow for the entropy to be provided by
\begin{equation}
\frac{\partial S}{\partial\lambda} = \frac{\partial I_G}{\partial S} + \eta S \,,
\end{equation}
which represents the reduced thermodynamic counterpart of the stochastic Ricci flow introduced in Sec.~\ref{sec:1}.

 
 We denote the value of the event horizon entropy as $S_{0}=\pi r_{H}^2$, and the potential $V(S,l)$ at $l_0$ as $V_0(S)$. The perturbed entropy can be expressed as $S=S_0+\beta$, with $\beta\ll 1$. The entropy perturbation can be then written as
\begin{equation}
     \tilde{S}=S-S_0=\beta\,.
 \end{equation}
Correspondingly, the stochastic gradient flow equation Eq.~\eqref{eq:free energy Ricci flow} can be recast as
\begin{equation}
     \frac{\partial \tilde{S}}{\partial \lambda}=-\frac{\partial \left(V-V_0\right)}{\partial \tilde{S}}+\eta(\lambda)\,\tilde{S} \,.\label{eq:free energy Ricci flow 3}  
 \end{equation}
This means that the entropy perturbation $\tilde{S}$ is driven by the thermodynamic potential difference $V-V_0$, determined by the multiplicative white noise.\\

\begin{figure}[htbp]
\centering
\includegraphics[width=.4\textwidth]{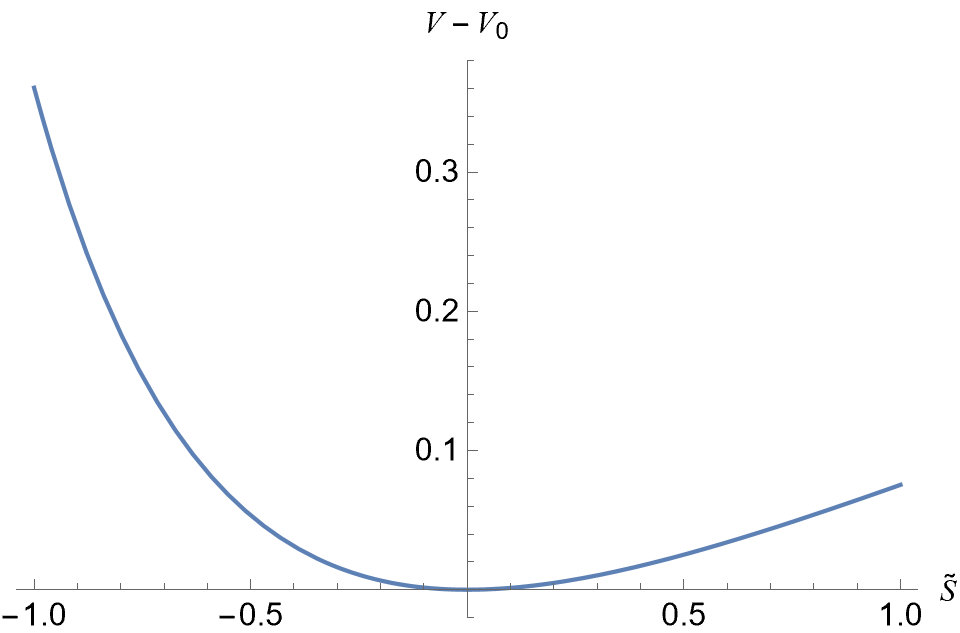}
\qquad
\includegraphics[width=0.4\textwidth]{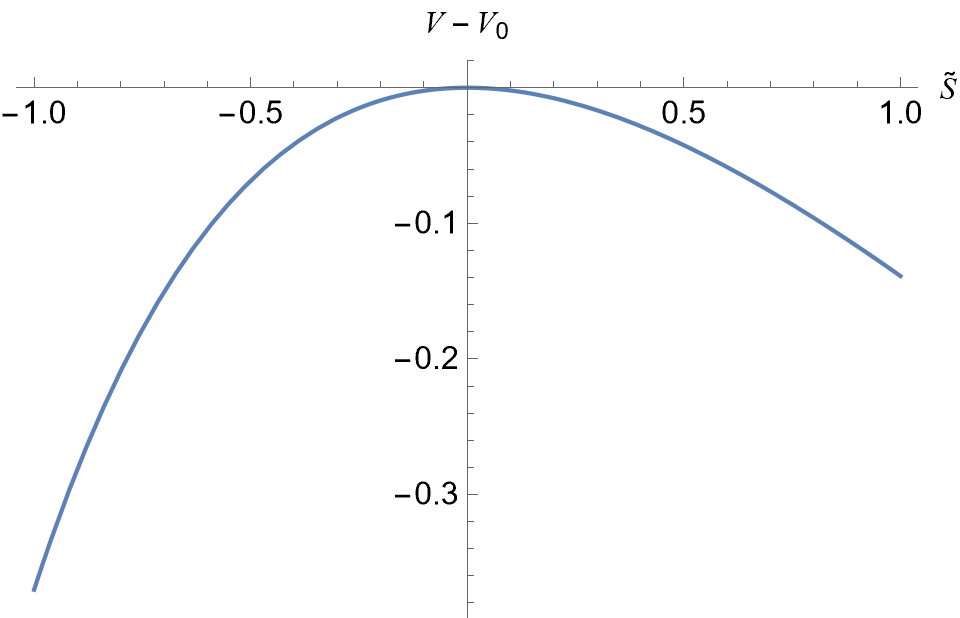}
\caption{We show the potential difference \( V - V_0 \) as a function $S$ for two cases: \( l^2 = 2.25 \) (upper panel) and \( l^2 = 6 \) (lower panel).}\label{fig_potential}
\end{figure}

In Fig.~\ref{fig_potential}, we plot the potential difference $V-V_0$ as a function of $S$ at $l^2=2.25$ and $l^2=6$. This figure indicates that any perturbation of $S$ tends to approach $0$ for $l^2=2.25$, when the A-dS radius lies inside the event horizon $l_0$, while it moves away from $0$ for $l^2=6$, when the A-dS radius lies outside the event horizon $l_0$. In the former case, the point $0$ is a local minimum of the potential difference, whereas in the latter case it becomes a maximum, signaling an unstable potential energy landscape. 

\subsection{Simulation result}
\begin{figure}
	\centering 
\includegraphics[width=0.45\textwidth, angle=0]{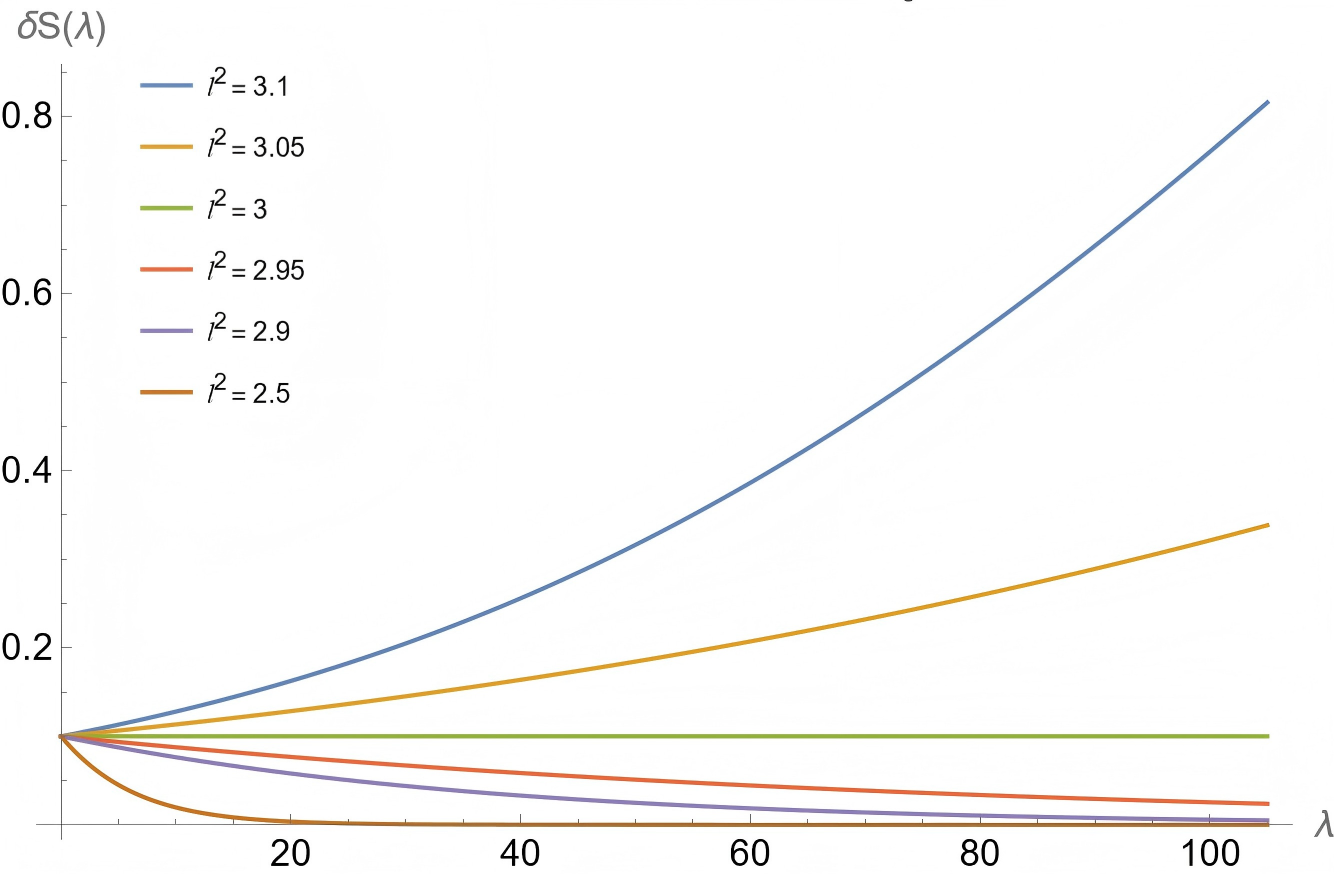}
	\caption{We display the evolution along the flow time $\lambda$, out of the S-AdS black hole equilibrium configuration, of the entropy perturbation $\delta S$, considering different values of $l^2$.} 
	\label{fig_0}
\end{figure}

\noindent
If we neglect the stochastic noise, from Eq.~\eqref{eq:free energy Ricci flow} and the shape of the potential difference in Fig.~\ref{fig_potential}, depending on the convexity or concavity of this curve, we can infer the evolution of the perturbation $\tilde{S}=\delta S$. When the curve has a convexity, the equilibrium point $\tilde{S}=0$ appears as a local minimum of the potential difference; correspondingly perturbations around $\tilde{S}$ decay, hence enabling to recover $\tilde{S}=0$. Differently, when the curve has a concavity, the same point becomes a maximum, implying a growth of the perturbations. The evolutions are displayed in Fig.~\ref{fig_0}, which shows the results of the numerical simulation. \\

From Fig.~\ref{fig_0}, we observe that for $l<l_0$ the perturbation of the horizon area dissipates as the flow time increases. In contrast, when $l>l_0$, the perturbation increases under the flow, leading to a thermodynamic instability. \\

Considering now the influence of the stochastic noise, besides the theoretical analysis of the stability properties, we may further  investigate numerically the dynamical behavior of the system. One possible approach is to perform Monte Carlo simulations over a large number of stochastic trajectories in order to study the statistical behavior of the entropy perturbations. Another approach is to solve the corresponding Fokker--Planck equation, which governs the evolution of the probability distribution associated with the stochastic process. 

\section{Monte Carlo and Fokker--Planck analysis} \label{MC}
\subsection{Monte Carlo simulation}
\begin{figure}
	\centering 
\includegraphics[width=0.45\textwidth, angle=0]{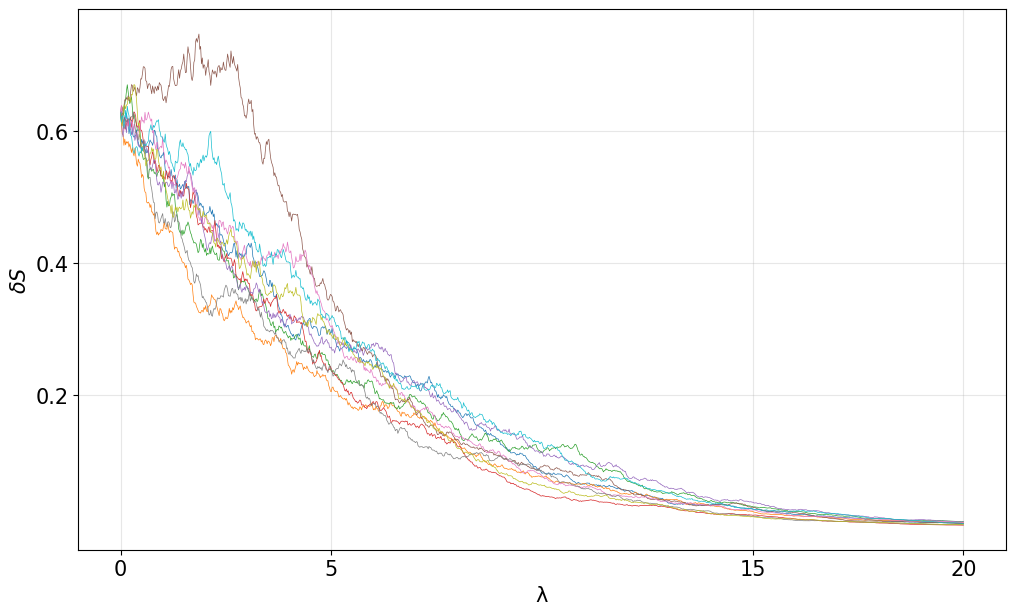}
	\caption{We display ten trajectories of the entropy perturbation $\delta S$ along the flow time $\lambda$, the AdS radius being taken to be $l^2=2.25$.} 
	\label{Mfig_1}
\end{figure}

\begin{figure}
	\centering 
\includegraphics[width=0.45\textwidth, angle=0]{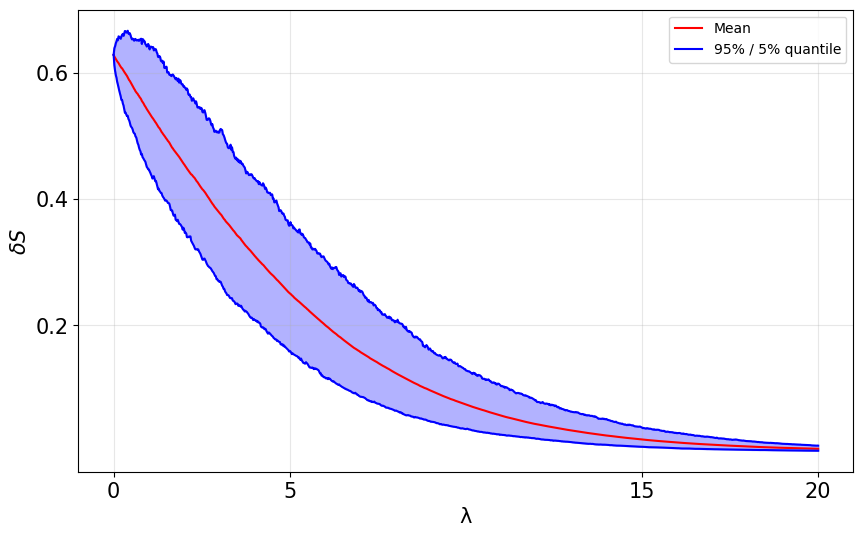}
	\caption{We display the mean trajectory and the quantile range of the entropy perturbation $\delta S$ with respect to the flow time $\lambda$, for the value of the AdS radius $l^2=2.25$.} 
	\label{Mfig_2}
\end{figure}

\noindent
The drift force term in the S-AdS case is calculated explicitly as
\begin{equation}
    \frac{\partial V}{\partial S}=\frac{1}{3}-\frac{4}{3\left(1+\frac{3S}{\pi l^2}\right)^2}\,.
\end{equation}
We then use the Euler-Maruyama finite difference method~\cite{Higham}. The stochastic evolution equations are then discretized according to the It\^o prescription, consistently with the interpretation adopted in Sec.~\ref{sec:1}. The gradient flow equation Eq.~\eqref{eq:free energy Ricci flow 2}, applying finite differences, recasts as
\begin{equation}
    \tilde{S}^{i+1}=    \tilde{S}^{i}+\left(\frac{4}{3\left(1+\frac{3\tilde{S}^{i}}{\pi l^2}\right)^2}-\frac{4}{3\left(1+\frac{3\tilde{S}^{i}}{\pi l_0^2}\right)^2}\right)\Delta \lambda +\tilde{S}^i \Delta W\,,
\end{equation}
where $\Delta W= W^i \sqrt{\Delta \lambda}$ represents the increment of the Wiener process over the interval $\Delta \lambda$, with $W^i$ being a random variable sampled from the standard Gaussian distribution $\mathcal{N}(0,1)$, the superscript index $i$ representing the value at the $i$-th step in the discretized flow time $\lambda$, namely  
\begin{equation}
    \lambda_{i}=i\triangle \lambda\,,
\end{equation}
\begin{equation}
    \tilde{S}_{i}=\tilde{S}(\lambda_{i})\,.
\end{equation}
Since the explicit expression for the drift function is known, the drift-force term $\frac{\partial V}{\partial S}$ does not generally require further discretization. We take the initial perturbation to be $\beta=0.2\pi$, and the stochastic strength parameter to be $\alpha=0.1$. For the case $l^2=2.25$, we sample the flow time using $500$ points with maximum flow time $\lambda_{max}=20$. The stochastic evolution is discretized using $N=1000$ steps, corresponding to the step size $\Delta \lambda=\frac{\lambda_{max}}{N}=0.02$. The resulting behavior is shown in Figs.~\ref{Mfig_1}-\ref{Mfig_2}, including ten stochastic trajectories together, with the mean trajectory, and the associated quantile bounds. We observe that, within this case $l^2=2.25$ with stochastic strength $\alpha=0.1$, the entropy perturbation decays to zero at a flow time corresponding to $\lambda\sim20$.

\vspace{0.2 cm}

\begin{figure}
	\centering 
\includegraphics[width=0.45\textwidth, angle=0]{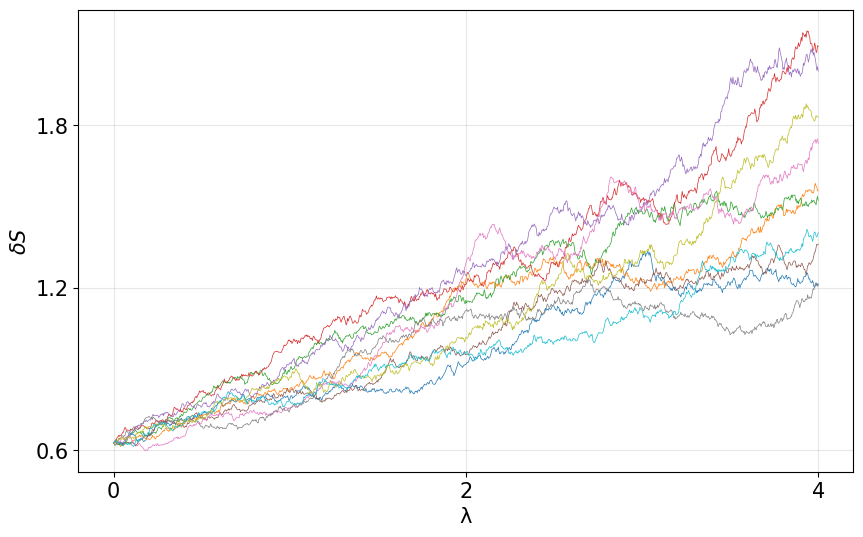}
	\caption{Ten trajectories of the entropy perturbation $\delta S$ with the flow time $\lambda$. The AdS radius $l^2=6$.}
	\label{Mfig_3}
\end{figure}

\begin{figure}
	\centering 
\includegraphics[width=0.45\textwidth, angle=0]{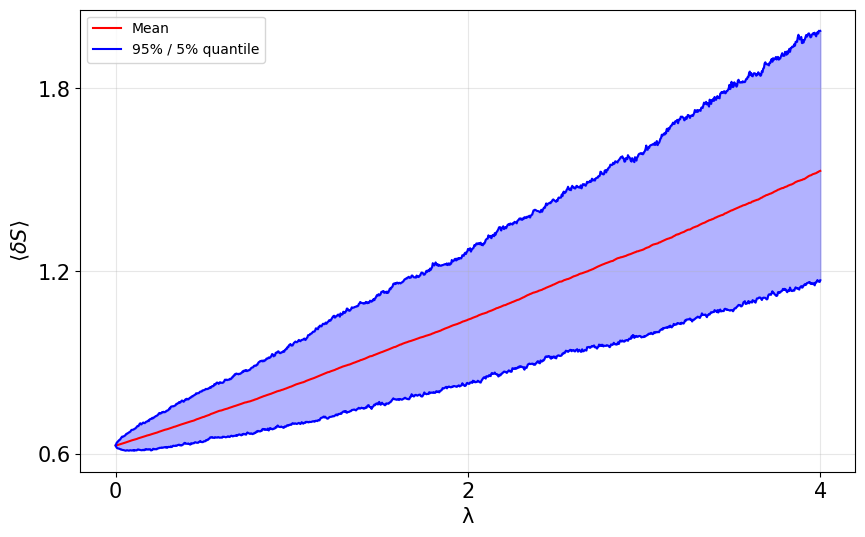}
	\caption{Mean trajectory and quantile range of the entropy perturbation $\delta S$ with respect to the flow time $\lambda$, for the AdS radius $l^2=6$.} 
	\label{Mfig_4}
\end{figure}

We then consider the case $l^2=6$. In this case, the flow time is sampled using $500$ points with maximum flow time $\lambda_{max}=4$. Once again we use $N=1000$ discretization steps, yielding $\Delta \lambda=\frac{\lambda_{max}}{N}=0.004$. The corresponding stochastic behavior is presented in Figs.~\ref{Mfig_3}-\ref{Mfig_4}, where we display ten sample trajectories, along with the mean trajectory and the quantile bounds. The numerical results show that, for the case $l^2=6$ with $\alpha=0.1$, the entropy perturbation growth rapidly and experiences increasingly large-amplitude fluctuations as flow time progresses.
\subsection{The Fokker-Planck equation}\label{Sec:FP}
\noindent
To explore the evolution and properties of a stochastic dynamical system, it is often useful to study its probability distribution. Although the random force generates different trajectories for each realization of the stochastic process, the corresponding probability distribution exhibits a well-defined collective behavior. Therefore, instead of focusing on individual trajectories, we turn to the study of the associated Fokker–Planck (FP) equation, which governs the evolution of the probability distribution of the system. Since the stochastic differential equation is interpreted in the It\^o sense, the corresponding probability density obeys the standard It\^o Fokker–Planck equation.\\

The FP equation related to Eq.~\eqref{eq:free energy Ricci flow 2} is
\begin{equation}
    \frac{\partial \rho(\lambda,S)}{\partial \lambda}=-\frac{\partial }{\partial S}\left( \rho \frac{\partial V}{\partial S}\right)+\frac{\alpha}{2}\frac{\partial^2}{\partial S^2}\left(\rho S^2\right)\,,
\end{equation}
and the corresponding formula for Eq.~\eqref{eq:free energy Ricci flow 3} is 
\begin{equation}
       \frac{\partial \rho(\lambda,\tilde{S})}{\partial \lambda}=-\frac{\partial }{\partial S}\left( \rho \frac{\partial (V-V_0)}{\partial S}\right)+\frac{\alpha}{2}\frac{\partial^2}{\partial S^2}\left(\rho \tilde{S}^2\right)\,.
\end{equation}
While numerically solving partial differential equations, appropriate boundary conditions must be imposed. Here, one can assume the probability density $ \rho$ and/or its derivative $\frac{\partial \rho}{\partial S}$ to vanish at the boundaries, the former one taking into account absorption, the latter one is reflection. We have then imposed absorbing boundary conditions. These boundary assumptions are not only used here for the numerical treatment of the Fokker–Planck equation, but also naturally enter the derivation of the Fokker–Planck equation from the underlying stochastic differential equation through the probability elimination of boundary terms~\cite{stochastic_1}, namely
\begin{equation}
    \begin{aligned}
&\rho(\lambda,\tilde{S}_{BC})=0\,,
    \end{aligned}
\end{equation}
\begin{equation}
    \begin{aligned}
&\frac{\partial \rho(\lambda,\tilde{S}_{BC})}{\partial \tilde{S}}=0\,,
    \end{aligned}
\end{equation}
where  $\tilde{S}_{BC}$ represents the boundaries $\tilde{S}_{max}$ and $\tilde{S}_{min}$, which are considered to be, respectively, the maximum and minimum values of entropy that are attainable in the stochastic process. We then consider 
\begin{equation}
    \tilde{S}_{min}=-0.3\,\pi, \qquad  \tilde{S}_{max}=0.3\,\pi\,,
\end{equation}
the initial distribution being
\begin{equation}
    \rho(\lambda=0,\tilde{S})=\frac{1}{0.1\sqrt{2\pi}}e^{-\frac{\left(\tilde{S}-\tilde{S}_{0}\right)^2}{2\times 0.1^2}}\,.
\end{equation}
This represents a Gaussian normal distribution for the initial entropy perturbation $\tilde{S}$ at the thermal/flow time \( \lambda = 0 \), with mean \( \tilde{S}_0 \) and standard deviation \( 0.1 \). This indicates that the initial entropy is narrowly concentrated around $ \tilde{S}_0$, with small fluctuations on the order of \( 0.1 \). We set $\tilde{S}_0=0.1\pi$, and use a spatial grid of 50 points, with flow time interval $\Delta \lambda=0.5 (\Delta S)^2\approx 0.0007$.\\ 

\begin{figure}
\centering 
\includegraphics[width=0.45\textwidth, angle=0]{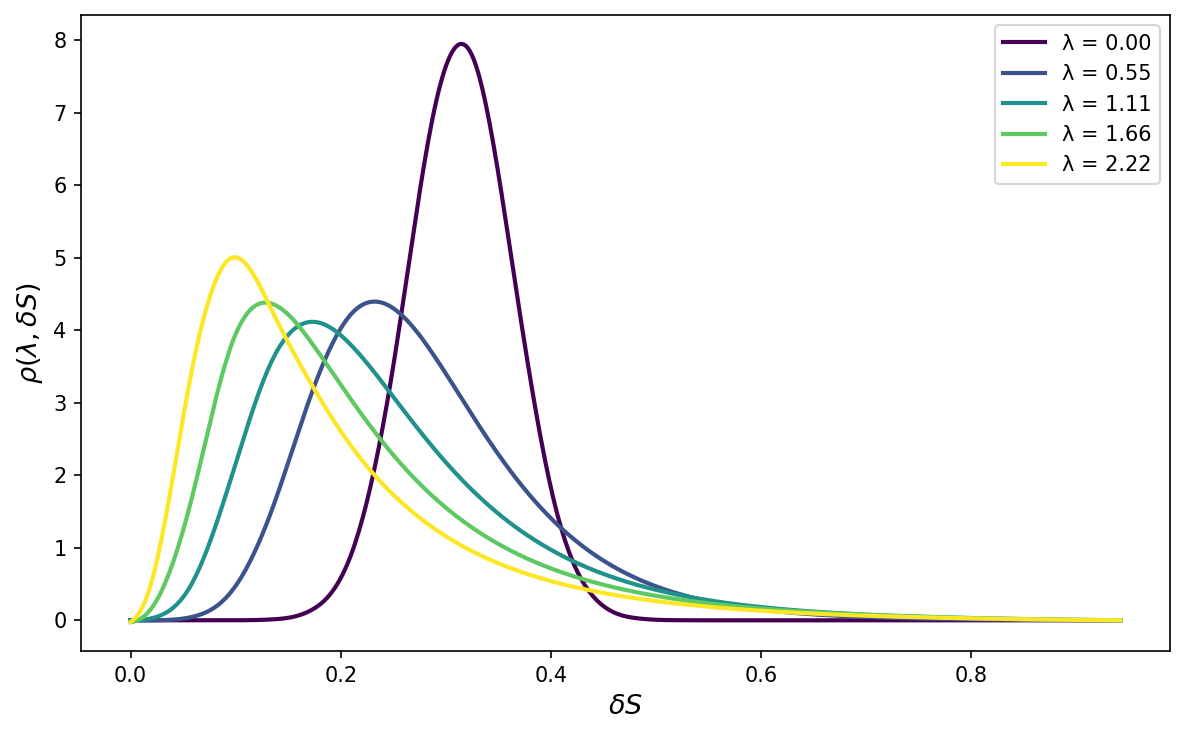}
	\caption{We display the probability $\rho(\delta S,\lambda)$ at different flow time, for $l^2=2.25$ and the stochasticity strength $\alpha=0.1$.}
	\label{fig:FP1}
\end{figure}

\begin{figure}[htbp]
\centering
\includegraphics[width=.45\textwidth]{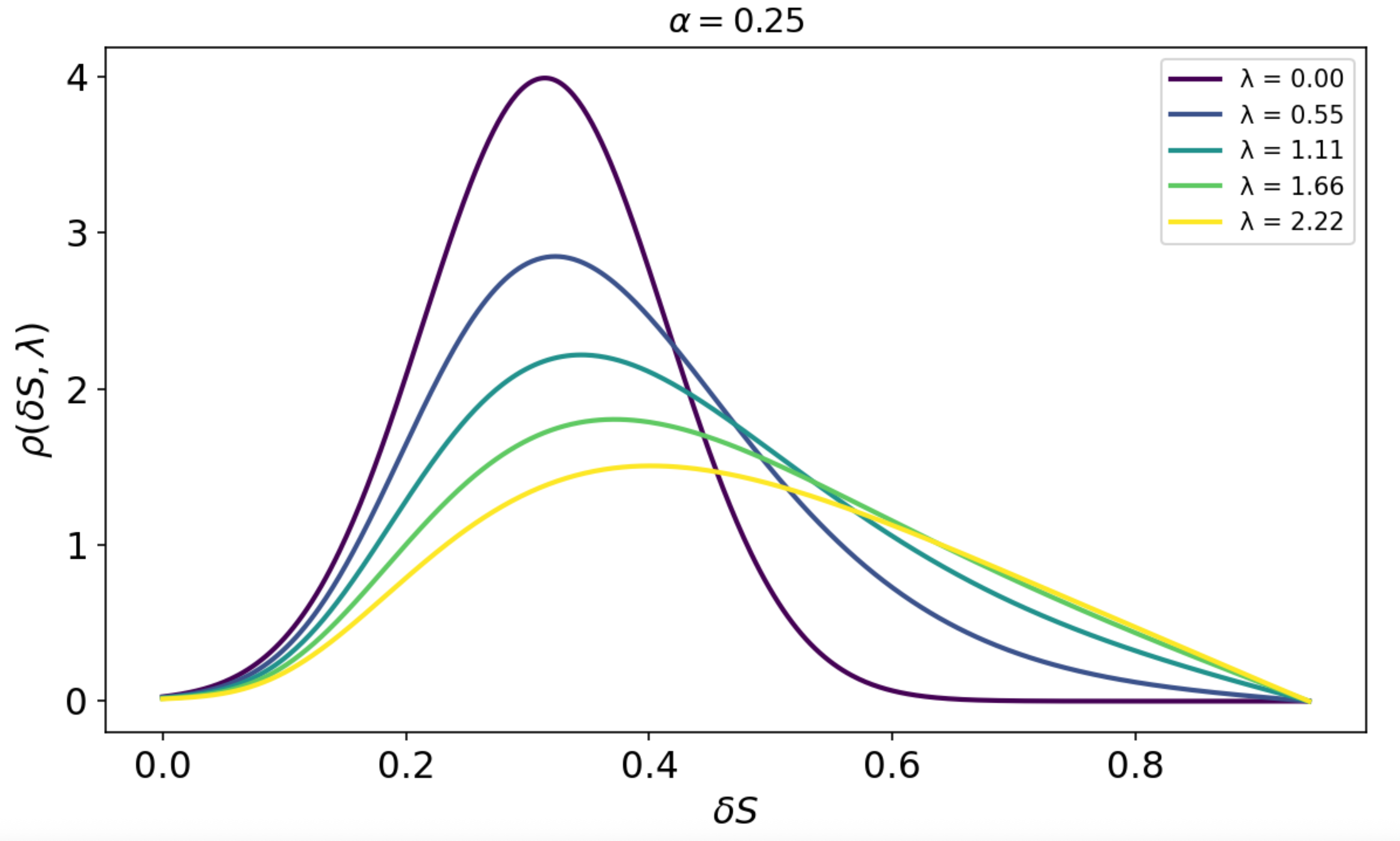}
\qquad
\includegraphics[width=0.45\textwidth]{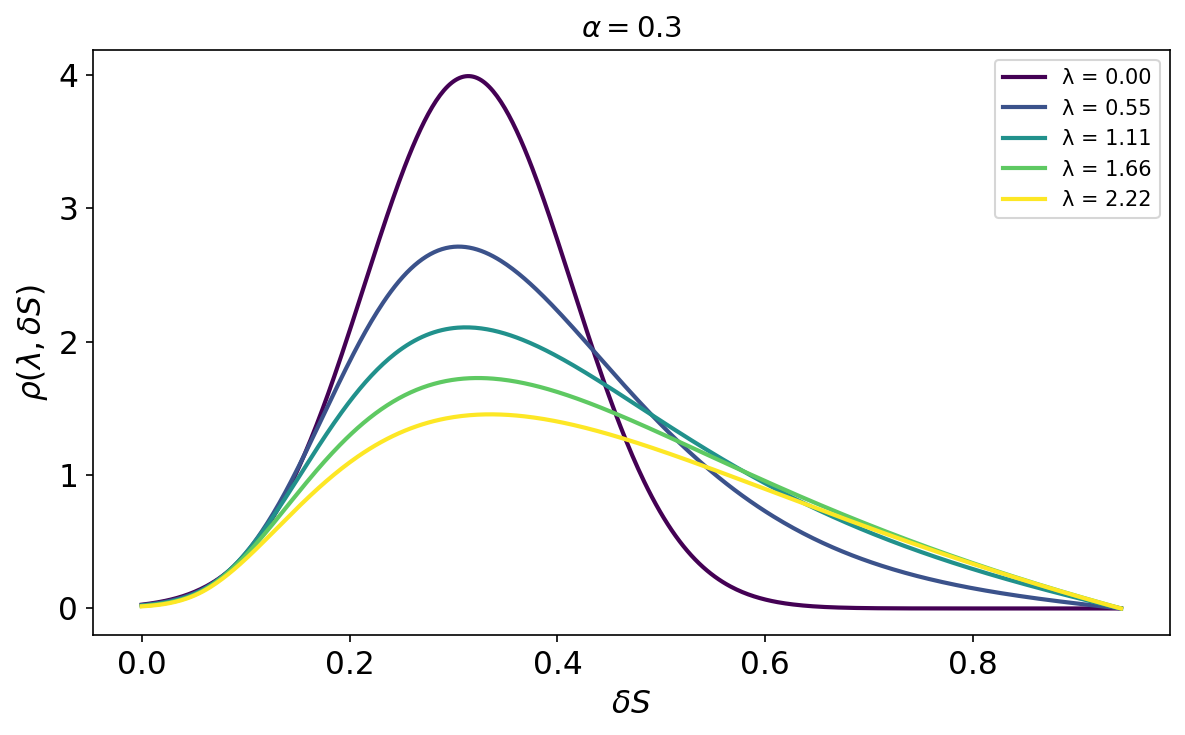}
\qquad
\includegraphics[width=0.45\textwidth]{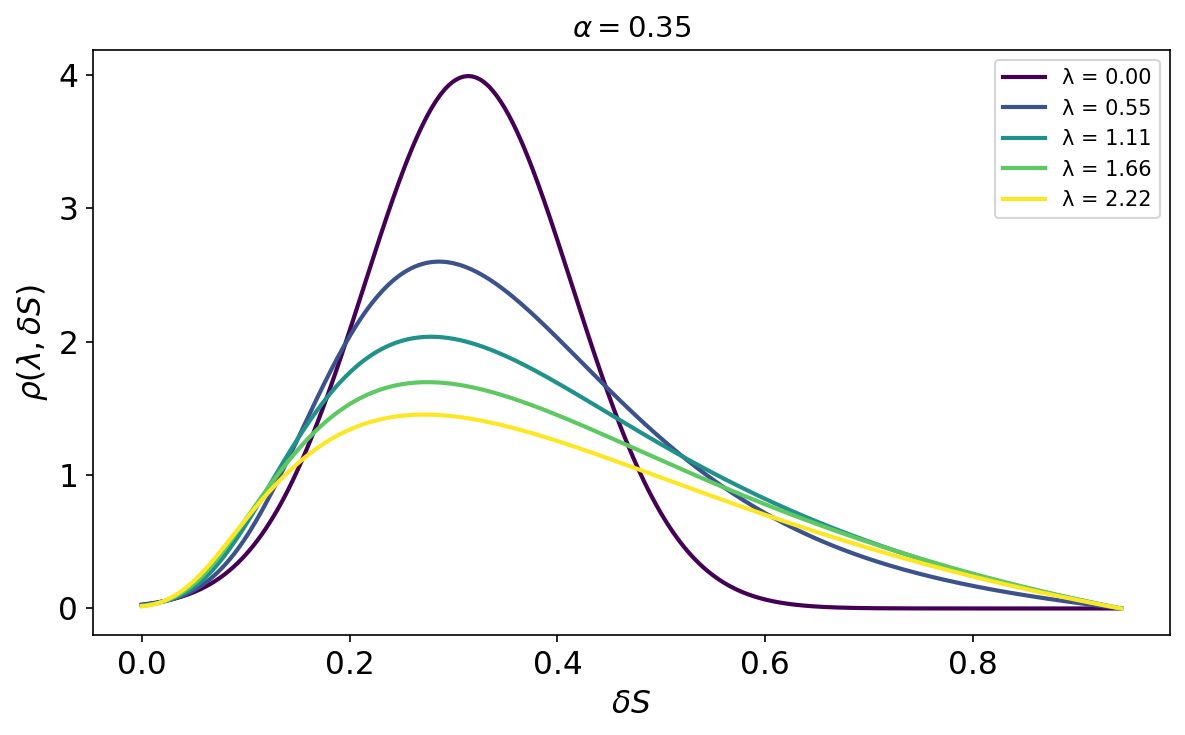}
\caption{We display the probability $\rho(\delta S,\lambda)$ at different flow time, for $l^2=6$. The panels from top to bottom correspond to $\alpha=0.25,\alpha=0.3,\alpha=0.35$, respectively. }\label{fig:FP2}
\end{figure}

We then consider two representative cases in our simulations. The first case, with  $l^2=2.5$, and the time-slice evolution of the probability distribution is shown in Fig.~\ref{fig:FP1}. The second case corresponds to $l^2=6$ , with corresponding probability distribution presented in Fig.~\ref{fig:FP2}. In Sec.~\ref{subsec:2} we observe that, after introducing stochasticity, the equilibrium configuration that was originally unstable for small cosmological constant is stabilized. To investigate this effect, we consider several different stochastic strengths, namely $\alpha=0.25$, $\alpha=0.3$ and $\alpha=0.35$.
As indicated by the thermodynamic potential analysis in Sec.~\ref{Sec:Thermo}, the case $l^2=6$ lies above the critical value, while the case $l^2=2.25$ lies beneath the critical value.\\

Comparing Fig.~\ref{fig:FP1} with Fig.~\ref{fig:FP2}, in the evolution of the density distribution for the $l^2=2.25$ case, we observe that the peak quickly shifts toward smaller values, while the variance of the probability distribution remains narrow, without extending toward larger values --- this indicates an overall decrease. In the second case $l^2=6$, however, although the peak shifts to smaller value when $\alpha>0.3$, the variance widens. This means that most trajectories shift to smaller value of $\delta S$, but some still increase. Overall, the distribution in this case does not shrink to a concentrated range, but instead decreases in a dispersed manner. When the stochasticity is weak, for instance $\alpha=0.25$, the evolution is still predominantly governed by the deterministic drift force, namely by the gradient of the thermodynamic potential, causing the entropy perturbation to grow with flow/thermal time. This behavior, as can be observed from Fig.~\ref{fig_M2}, is consistent with results obtained from the stochastic Ricci flow analysis of the perturbed metric in Sec.~\ref{sec:2}. When the stochastic strength becomes sufficiently large, such as in the case $\alpha=0.35$, the perturbation instead exhibits a dissipative tendency. This suggests that strong stochastic effects can effectively suppress the growth of perturbations and induce qualitatively a restorative behavior of previous equilibrium configurations. Physically, this implies that entropy perturbations --- regardless of whether they originate from horizon deformations, surface growth, or changes induced by infalling matter --- tend to relax over thermal time. Remarkably, this recovery behavior is observed not only in the positive heat-capacity region, which is conventionally regarded as thermodynamically stable, but also extends into the negative heat-capacity region, where the system would ordinarily be expected to be thermodynamically unstable.

\section{Summary and conclusions}\label{sec:3}
\noindent
We have studied the Ricci flow complemented with stochastic noise, understood as a Langevin-type diffusion flow of the gravitational action, in order to investigate the thermodynamic stability of the equilibrium Schwarzschild-AdS (S-AdS) solution of the Einstein equations. The numerical Ricci (target) flow is applied to the spherically perturbed S-AdS metric, the cosmological term being treated as matter entering the Ricci target term. We relied on the numerical framework and the code delivered by Headrick and Wiseman while discussing the Schwarzschild metric \cite{RF2}, and extended their results to the S-AdS solution, featuring a heat capacity divergence point. The metric functions evolve under the Ricci flow keeping the restricted form: the $tt$ and $rr$ components are assumed to retain their original forms, whereas the spherical part component contains an initially increased perturbation. The numerical results show that the perturbed part will converge to the S-AdS solution for sufficient large absolute values of $\Lambda$, while will expand for smaller value of $\Lambda$. Furthermore, by introducing various strength of the multiplicative noise, which may be interpreted as arising from the Parisi–Wu \cite{Parisi:1981fw} stochastic quantization framework. The stochastic simulations exhibit a behavior correlated with the heat-capacity divergence point. However, in contrast to the purely deterministic case, we find that stochasticity plays a stabilizing role in the evolution of the system.\\

In Ref.~\cite{Li_Wang_1}, then authors have used an extended off‑shell Gibbs free energy, which changes the temperature from the standard black hole temperature, and considered a gradient thermal flow equation --- a Langevin‑type equation. In our analysis, we have followed a similar approach but taken the black hole free energy to be multiplied by the inverse temperature, as the thermodynamical potential. Additionally, we have considered perturbations in the entropy $S$, as well as the potential difference between the value of $l$ and its value at the heat‑capacity divergence point  $l_0$. We have then performed a standard numerical Monte Carlo simulation and, to achieve a more detailed analysis of probability evolution, solved the Fokker‑Planck (FP) equation. This phenomenological framework has proved to be capable of reproducing the qualitative features of the (stochastic) Ricci flow dynamics. In this regard, the approach we have followed may provide a useful reference point. \\

For $l^2=6$, the interpretation is more subtle. When the stochastic strength is sufficiently large, the peak of the probability distribution shifts toward smaller entropy perturbations, indicating that stochasticity suppresses the deterministic runaway growth. However, the width of the distribution also increases, showing that stochastic excursions remain sizable. Thus, in the negative heat-capacity branch, stochasticity should not be interpreted as producing tight concentration around the equilibrium configuration, as in the $l^2=2.25$ case. Rather, it provides an effective suppression of the instability and prevents the deterministic growth of the perturbation from dominating the dynamics.\\

Our analysis has implemented a stochastic Ricci flow dynamics in the flow/thermal time of the black hole metric, with a special focus on the problem of negative heat capacity, and on the heat capacity divergence point in black hole physics. The negative heat capacity is a very special phenomenon that occurs in gravitational systems. As we know from astrophysics, stellar systems exhibit negative heat capacity. This contradicts general thermodynamic interactions, which tend to homogenize different stars and drive them toward a uniform distribution --- whereas gravity intrinsically draws them together. Such an interaction is fundamentally different from ordinary thermal behavior. On the other hand, negative heat capacity also indicates that black hole systems, as well as gravitational black holes themselves, are not thermodynamically stable. However, while using the stochastic Ricci flow and the stochastic gradient flow of the entropy $S$, we have shown that for the Schwarzschild-AdS black hole, which possesses thermodynamically stable and unstable branches separated by the Davies heat-capacity divergence point, the introduction of stochasticity leads to a different conclusion. Specifically, a strong Brownian motion in the thermal background can actually help stabilize the black hole: it suppresses the growth of entropy perturbations and even dissipates them.

\section*{Acknowledgements}
\noindent 
We thank Ugo Moschella for valuable discussions and insightful suggestions during the development of this work.

\appendix

\bibliographystyle{unsrt} 
\bibliography{reference}

\end{document}